# A Utility Based Approach to Energy Hedging

**John Cotter[a] and Jim Hanly[b]**

November 2010

[a]John Cotter,
Director of Centre for Financial Markets,
School of Business
University College Dublin,
Blackrock,
Co. Dublin,
Ireland,
Tel +353 1 716 8900,
e-mail john.cotter@ucd.ie.
Corresponding Author

[b]Jim Hanly,
School of Accounting and Finance,
Dublin Institute of Technology,
Dublin 2,
Ireland.
tel +353 1 402 3180,
e-mail james.hanly@dit.ie.

This research was conducted with the financial support of Science Foundation Ireland under Grant Number 08/SRC/FM1389.

## A Utility Based Approach to Energy Hedging

### Abstract

A key issue in the estimation of energy hedges is the hedgers' attitude towards risk which is encapsulated in the form of the hedgers' utility function. However, the literature typically uses only one form of utility function such as the quadratic when estimating hedges. This paper addresses this issue by estimating and applying energy market based risk aversion to commonly applied utility functions including log, exponential and quadratic, and we incorporate these in our hedging frameworks. We find significant differences in the optimal hedge strategies based on the utility function chosen.

## 1. Introduction

Optimising hedging strategies for energy products such as Oil and Natural Gas is a key issue for energy hedgers given the importance of these products within the global economy and because of their susceptibility to price volatility (Regnier, 2007). The risk attitude of hedgers as expressed by their utility function has an important role to play in the determination of what is considered optimal from a hedging perspective. While many papers have looked at optimal hedging (Kroner and Sultan 1993, Cotter and Hanly, 2006) and some have looked specifically at energy hedging (Chen, Sears and Tzang, 1987), to the best of our knowledge none have focused on the impact of differing risk attitudes on optimal energy hedging strategies, Also, the literature has not tended to explicitly model risk aversion. Instead either infinite risk aversion is assumed, or arbitrary values are used to estimate optimal ratios (OHR's). Furthermore, little attention has been paid to the need to allow risk aversion to vary over time in response to changes in attitudes towards risk. These issues are more pertinent than ever given the recent turmoil in financial markets which has illustrated how investor perceptions towards risk can change.

This paper addresses these issues and contributes to the literature in a number of ways. Firstly, we estimate and apply a time varying risk aversion coefficient using an approach that focuses on energy market participants. Secondly, we apply the resulting risk aversion to estimate optimal hedging strategies for two of the most important energy assets, namely Crude Oil and Natural Gas, using three different utility functions, the quadratic, exponential and log, to allow us to examine how differing risk attitudes will affect the determination of the OHR. These utility functions cover a variety of risk attitudes that are commonly attributed to economic agents (Alexander, 2008), and specifically in hedging (Ederington, 1979; Cecchetti, Cumby and Figlewskim 1988; and Brooks, Cerny and Miffre, 2007). Thirdly we compare the utility based hedges with a minimum variance hedge ratio (MVHR). For each of the different hedges we use a multivariate GARCH model to estimate the underlying variance covariance matrix. This allows us to compare OHR's on the basis of different utilities rather than on the basis of different approaches to estimating the underlying variance covariance matrix. Finally we examine our utility based hedges in an out-of-sample setting using a unique approach that allows us to incorporate risk aversion into our forecasted OHR's.

Our results show significant differences between each of the utility based OHR's particularly where there is skewness and kurtosis in the data. Since these characteristics are typical of energy price data, this indicates the importance of specifying a utility function that reflects the risk attitude of energy hedgers. We also find that the risk preferences of energy hedgers tend to vary over time and are similar to those reported in the broader asset pricing literature for equity investors (see, for example, Ghysels, Santa Clara and Valkanov, 2005). This highlights the need to explicitly model risk aversion that changes over time rather than arbitrarily applying a single value for hedging purposes.

The remainder of this paper proceeds as follows. Section 2 details the optimal hedging framework and the role of risk aversion in the estimation of optimal hedging strategies. In Section 3 we outline the different utility functions used together with the coefficient of relative risk aversion (CRRA). Section 4 describes the estimation procedure for both the risk aversion parameter and the different utility based hedges. The data is detailed in section 5; empirical results are presented in section 6 and concluding remarks in section 7.

## 2. Utility and Risk Preferences

In this paper we estimate and compare hedging strategies for hedgers with different attitudes towards risk as defined by their utility functions. We now discuss the three different utility functions we use to characterize energy hedgers.

### 2.1 Risk Preferences and the CRRA

The utility function and risk aversion of a hedger reflects their view of the tradeoff between risk and return. There are two different characterizations of risk aversion. Absolute risk aversion (ARA) is a measure of hedger reaction to dollar changes in wealth. This is the relative change in the slope function at a particular point in their utility curve[1]. The CRRA differs from the ARA in that it examines changes in the relative

[1] This refers to assumptions re changes in risk preferences as wealth changes. To measure an hedger's absolute risk aversion we use $\frac{-U''(Wealth)}{U'(Wealth)}$

percentages invested in risky and risk free assets as wealth changes. We define the CRRA as follows:

$$\text{CRRA} = -W * \frac{U''(Wealth)}{U'(Wealth)} \tag{1}$$

where $W$ is the wealth of the investor. The CRRA and ARA are broadly similar although the CRRA has a scaling factor to reflect the investor's current level of wealth (Arrow, 1971). We use the CRRA to capture the hedger's attitude towards risk in a single number which we can use to tailor the hedging strategies to the risk and return preferences of hedgers with different utility functions. We view the CRRA within its role as a determinant of the market risk premium.

## 2.2 Quadratic Utility

The Quadratic Utility function is one of the most applied in finance and economics in such areas as portfolio theory, asset pricing and hedging (Danthine and Donaldson, 2003). In portfolio theory, quadratic utility was instrumental in the development of the minimum variance portfolio (Markowitz, 1952). Similarly in the hedging literature it is a key assumption in much of the literature on optimal hedging (Ederington, 1979). It is defined as follows:

$$U(W) = W - aW^2, \qquad a > 0 \tag{2a}$$

Define U ( ) as the utility function and W as wealth, a is a positive scalar parameter measuring risk aversion. The first and second derivatives of this are given by:

$$U'(W) = 1 - 2aW \tag{2b}$$

$$U''(W) = -2aW \tag{2c}$$

To be consistent with non-satiation where utility is an increasing function of wealth implying more is preferable to less, the following restriction is placed on W:

$$U'(W) = 1 - 2aW > 0$$

The relative risk aversion measure is:

$$R(W) = \frac{2aW}{1-2aW} \tag{2d}$$

The quadratic utility function is consistent with a hedger who decreases the dollar amount invested in risky assets as wealth increases. This means that the proportion invested in risky assets will decrease as wealth increases. This implies increasing relative risk aversion. An agent with quadratic utility and whose wealth has increased may no longer need to target riskier asset classes as they no longer wish to make higher returns.

## 2.3 The Log Utility Function

The log utility function has been frequently used since Bernoulli (1738) first promoted the concept of utility. In the hedging literature, Cecchetti, Cumby and Figlewski (1988) compared optimal hedges for investors with log utility with those for investors with Quadratic utility. Log utility is defined as follows:

$$U(W) = \ln W \tag{3a}$$

$$U'(W) = W^{-1} \tag{3b}$$

$$U''(W) = -W^{-2} \tag{3c}$$

For an hedger with log utility, the relative risk aversion measure is:

$$R(W) = \frac{-W(-W^{-2})}{W^{-1}} = 1 \tag{3d}$$

Thus the log utility function is consistent with constant relative risk aversion which means the proportion invested in risky assets will remain constant for all levels of wealth. This type of utility function would suit an investor who takes the view that risk and wealth are independent and who tailors their investment strategy accordingly.

## 2.4 The Exponential Utility Function

The last of our utility functions, exponential utility has been broadly applied with a number of studies finding that it is a reasonable representation of investor behavior (see for example, Townsend, 1994). It has also been applied in a hedging context by a number of papers including, Brooks, Cerny and Miffre (2007). It is defined as follows:

$$U(W) = -e^{-aW}, a > 0 \tag{4a}$$

$$U'(W) = ae^{-a*W} \tag{4b}$$

$$U''(W) = -a^2 e^{-a*W} \tag{4c}$$

$$R(W) = \frac{-W(-a^2 e^{-aW})}{ae^{-aW}} = -Wa \tag{4d}$$

Investors with Exponential utility invest constant dollar amounts and decreasing proportional amounts in risky assets as their wealth increases. Therefore this utility function is consistent with increasing relative risk aversion. This characterization of risk aversion is compatible with the idea that the wealthier an investor is, the more averse they will be to losses, although not as risk averse as those with quadratic utility.

## 3. Hedging and Risk Aversion

In this paper we approach the hedging problem from the perspective of an energy hedger who wishes to maximize their utility, where utility is a function of both risk and expected return. This allows us to incorporate the hedgers risk aversion into the choice of hedging strategy [2]. We choose three of the most well known [3] and applied characterizations of investor utility to examine how different attitudes towards risk will impact the hedging choice. These are Quadratic, Log and Exponential utility.

### 3.1 Definition of Hedgers

We look at hedging from the perspective of both short hedgers and long hedgers. Within an energy hedging setting, short hedgers are long the asset and are concerned with price decreases, whereas the long hedger is short the asset and is concerned with price increases. Therefore, their hedging outcomes will relate to opposite sides of the return distribution.

### 3.2 Optimal Hedge Ratio's

[2] An alternative approach is to choose an approach that minimizes the variance; however, this assumes that an investor is infinitely risk averse. For example, an investor who is infinitely risk averse would not consider investments which offered very large potential returns if it also meant taking on a small amount of additional risk. For more details on the specifics surrounding alternative approaches see Cotter and Hanly (2010).

[3] See for example, Merton (1990), Alexander (2008)

An optimal hedging strategy for both short and long hedgers can be derived as follows. Assuming a fixed spot position let $r_{st}$ and $r_{ft}$ be logarithmic returns on the spot and futures series respectively, and $\beta$ be the Optimal Hedge Ratio (OHR). The return to the hedged portfolio is constructed as follows:

$$R_p = +r_s - \beta r_f \qquad \text{(short hedger)} \qquad (5a)$$

$$R_p = -r_s + \beta r_f \qquad \text{(long hedger)} \qquad (5b)$$

The OHR $\beta$ is the weight of the futures asset in the hedged portfolio that is chosen to maximize expected utility. The OHR will differ depending on the utility function specified and the risk aversion of the hedger. In this paper we optimise the OHR using the three different utility functions, namely, Quadratic, Exponential and Log Utility. We also calculate the minimum variance hedge ratio (MVHR) for comparison purposes. This is a special case of the quadratic utility function where risk aversion is assumed to be infinite. We now outline the hedging models for each of the aforementioned utilities.

Assuming that the agent has a quadratic utility function, then the OHR can be calculated as:

$$\beta = \frac{-E(r_{ft})}{2\lambda\sigma_{ft}^2} + \frac{\sigma_{sft}}{\sigma_{ft}^2} \qquad (6)$$

where $E\ (r_{ft})$ is the expected return on futures, $\lambda$ is the risk aversion parameter , $\sigma_{ft}^2$ is the futures variance and $\sigma_{sft}$ is the covariance between spot and futures. The first term is speculative and the second term is the pure hedging or risk minimizing term. Equation (2) thus explicitly establishes the relationship between the risk aversion parameter $\lambda$ and the OHR. As risk aversion increases, the first term becomes smaller, such that for extremely large levels of risk aversion, the first term will approach zero. For the minimum variance hedge, since the objective is to minimize risk irrespective of return, we assume infinite risk aversion. The MVHR is therefore calculated using:

$$\beta = \frac{\sigma_{sft}}{\sigma_{ft}^2} \tag{7}$$

For an investor with log utility, relative risk aversion is constant and equal to 1, therefore setting $\lambda = 1$, we obtain the OHR for an investor with log utility;

$$\beta = \frac{-E(r_{ft})}{2\sigma_{ft}^2} + \frac{\sigma_{sft}}{\sigma_{ft}^2} \tag{8}$$

For an investor with exponential utility, their optimal hedge ratio will approximate the quadratic hedge for returns with normal distributions however for non-normally distributed returns; both skewness and kurtosis will have an impact. There is no simple closed form solution to this problem (for details see Alexander, 2008); however the optimal exponential hedge can be estimated by choosing $\beta$ to maximize the following expression:

$$\beta = \mu_{pt} - \frac{1}{2}\lambda\sigma_{pt}{}^{2} + \frac{\tau}{6}\lambda^2\sigma_{pt}^3 - \frac{\kappa - 3}{24}\lambda^3\sigma_{pt}^4 \tag{9}$$

where $\mu_{pt}$ is the expected return on the hedged portfolio, $\lambda$ is the risk aversion parameter, $\sigma^2_{pt}$ is the variance of the portfolio, $\tau$ is the skewness of the portfolio and $\kappa$ is the kurtosis. In general, for a long hedger with exponential utility, aversion to risk is associated with negative skewness and increasing variance and kurtosis.

## 4. Hedge Ratio Estimation

To obtain the OHR's for each of the different utility functions we require estimates of the risk aversion parameter together with the variance and covariances of the spot and futures for both Natural Gas and Oil. Estimation of the CRRA is based on the market risk premium for energy market participants, which is the excess return on a portfolio of assets that is required to compensate for systematic risk[4]. Within the asset pricing framework[5], the size of the risk premium of the market portfolio is determined by the aggregate risk aversion of investors and by the volatility of the market return as expressed by the variance.

$$E(r_{pt}) - rf = \lambda \sigma^2{}_{pt} \tag{10}$$

where $E(r_{pt}) - rf$ is the excess return on the market portfolio (or risk premium), $\lambda$ is the CRRA and ${\sigma_{pt}}^2$ is the variance of the return on the market. Intuitively, the CRRA depends on the size of the risk premium associated with a given investment.

[4] We use an energy index to proxy for the market to obtain risk aversion of energy hedgers. See Hanly and Cotter (2010) for a more comprehensive derivation of the CRRA.

[5] See Giovannini and Jorion (1989) for more details.

Consequently, the CRRA is the risk premium per unit risk (Merton, 1980). This general return volatility framework can be adjusted to account for any portfolio of assets. In this paper we use the DJ Stoxx Oil and Gas producers' index[6] as the market, as this consists of a broad range of companies involved in the production and supply of oil products, it should provide a good representation of the risk and return characteristics of the energy market. Therefore the risk aversion estimates we obtain will represent energy market participants rather than broader stock market participants as a whole. Defining $R_{pt} - \varepsilon_t = E(r_{pt}) - rf$ and setting $rf = 0$ [7], the adjusted equation can be written as:

$$R_{pt} = \lambda \sigma_{pt}^2 + \varepsilon_t \tag{11}$$

where $R_{pt}$ is the return on the hedged portfolio $\lambda$ is the CRRA and ${\sigma_{pt}}^2$ is the variance of the hedged portfolio and $\varepsilon_t$ is the error term.

## 4.1 CRRA Estimation

To estimate the CRRA, we use a GARCH-M specification (Engle et al, 1987) of the Diagonal Vech GARCH model of Bollerslev Engle and Wooldridge (1988). This model adjusted the mean equation to take account of the conditional variance of returns which links risk with reward[8]. By modelling the conditional mean and variance simultaneously we are able to obtain the CRRA in an efficient manner.

[6] Further details are at http://www.stoxx.com/download/indices/rulebooks/stoxx_indexguide.pdf
[7] These hedged portfolios do not include a risk free asset which we have assumed to be zero. For hedgers as distinct from investors, it is standard to assume that compensation for the risk free rate in an investment scenario may not be appropriate. The portfolios consist of just two assets, the unleaded spot and futures for each of the two energy assets we examine.

$$r_{pt} = \lambda \sigma_{pt}^2 + \varepsilon_t \quad (12)$$

$$[\varepsilon_t] | \Omega_{t-1} \sim N(0, \sigma_{pt}^2) \quad (13)$$

$$\sigma_{pt}^2 = \omega + \alpha \varepsilon_{t-1}^2 + \beta \sigma_{pt-1}^2 \quad (14)$$

where $r_{pt}$ is the return on the hedged portfolio, $\varepsilon_t$ is the residual, $\sigma_{pt}^2$ denotes the variance of the hedged portfolio, $\lambda$ is the CRRA, and $\Omega_{t-1}$ is the information set at time $t-1$, and $\omega, \alpha, \beta$ are parameters on the variance specification for the constant, lagged residuals and lagged variance respectively.

## 4.2 Hedging Model

We estimate the variance covariance matrix for spot and futures using a multivariate Diagonal Vech GARCH (1, 1) model. This is specified as follows:

$$r_{st} = \mu_{st} + \varepsilon_{st} \quad r_{ft} = \mu_{ft} + \varepsilon_{ft}, \qquad \begin{bmatrix} \varepsilon_{st} \\ \varepsilon_{ft} \end{bmatrix} | \Omega_{t-1} \sim N(0, \sigma_t) \quad (15)$$

$$\sigma^2{}_{st} = \omega_1 + \sum_{j=1}^{m} \alpha_{s,j} \varepsilon_{st-j}^2 + \sum_{k=1}^{n} \beta_{s,k} \sigma_{st-k}^2 \quad (16)$$

$$\sigma^2{}_{ft} = \omega_2 + \sum_{j=1}^{m} \alpha_{f,j} \varepsilon_{ft-j}^2 + \sum_{k=1}^{n} \beta_{f,k} \sigma_{ft-k}^2 \quad (17)$$

---

[8] They originally used an ARCH-M specification however it is more usual to use a GARCH-M specification given the advantages of the GARCH model over the ARCH.

$$\sigma_{sf_t} = \omega_3 + \sum_{j=1}^{m} \alpha_{sf,j} \varepsilon_{s\,t-j} \varepsilon_{f\,t-j} + \sum_{k=1}^{n} \beta_{sf,k} \sigma_{sf\,t-k} \qquad (18)$$

where $\Omega_{t-1}$ is the information set at time $t-1$, $\varepsilon_{st}, \varepsilon_{ft}$ are the residuals, $\sigma_{st}, \sigma_{ft}$ denotes the standard deviation of cash and futures and $\sigma_{sft}$ is the covariance between them. $\omega = (\omega_1, \omega_2, \omega_3)$ is a 3x1 vector, and $\alpha_j = (\alpha_{s,j}, \alpha_{f,j}, \alpha_{sf,j})$ and $\beta_k = (\beta_{s,k}, \beta_{f,k}, \beta_{sf,k})$ are 3x1 vectors. The model contains 3 +3m+3n parameters. The matrices $\alpha_j$ and $\beta_k$ are restricted to be diagonal. This means that the conditional variance of the cash returns depends only on past values of itself and past values of the squared innovations in the cash returns. The conditional variance of the futures returns and the conditional covariance between cash and futures returns have similar structures. Because of the diagonal restriction we use only the upper triangular portion of the variance covariance matrix, the model is therefore parsimonious, with only nine parameters in the conditional variance-covariance structure of the Diagonal VECH (1,1) model to be estimated.

### 4.3 PERFORMANCE OF HEDGING MODELS

We compare hedging performance using two different metrics, the variance and the Value at Risk (VaR). For the variance risk metric we use the percentage reduction in the variance of the cash (unhedged) position as compared to the variance of the hedged portfolio. This is given as:

$$HE_1 = 1 - \left[ \frac{VARIANCE_{HedgedPortfolio}}{VARIANCE_{UnhedgedPortfolio}} \right] \qquad (19)$$

This measure of effectiveness has been broadly applied in the literature on hedging (see Ederington, 1979) and is easy to understand and apply. The second hedging effectiveness metric is VaR. For a portfolio this is the loss level over a certain period that will not be exceeded with a specified probability. The VaR at the confidence level $\alpha$ is

$$VaR_{\alpha} = q_{\alpha} \tag{20}$$

where $q_{\alpha}$ is the relevant quantile of the loss distribution. The performance metric employed is the percentage reduction in the VaR of the hedged as compared with the unhedged position.

$$HE_2 = 1 - \left[ \frac{VaR_{1\% HedgedPortfolio}}{VaR_{1\% UnhedgedPortfolio}} \right] \tag{21}$$

The VaR is a useful risk metric risk given that it measures the potential money loss on a portfolio as well as a probability. Of more importance, the VaR allows us to examine the risk reduction for long and short hedgers separately, whereas in comparison the variance, assumes distribution symmetry makes no distinction on losses that may be experienced for long versus short hedgers. It has also been broadly applied as a measure of investor risk (see for example, Cabedo and Moya, 2003).

## 5. DATA AND ESTIMATION

The energy markets are a natural place to examine hedging given their overall importance in economic terms and also because of their susceptibility to supply and demand shocks together with the associated volatility. To estimate risk aversion for energy market participants we required a market index that was representative of the risk and return characteristics of the energy sector and that was broad enough to encompass the myriad different firms employed in this area. We therefore chose the Dow Jones Stoxx Oil and Gas Producers Index (Sector 0530) which covers companies involved in the exploration, production, refining, distribution and retail sales of Oil and Gas products. We also calculated the risk aversion coefficient using a variety of different energy indices which yielded very similar results, but our reported results are based on the DJ Stoxx index as it should provide us with a good indication of the risk and return characteristics of energy hedgers given the broad makeup of the index[9].

We select two key energy contracts, Oil and Natural Gas to examine energy hedging. For the Oil price we use the WTI Light Sweet Crude contract from CMEGROUP as it is a key international benchmark for oil pricing[10]. For Natural Gas we use the Henry Hub contract which is the primary natural gas setting price for North America. It also trades on CMEGROUP[11]. Both of these contracts were chosen as together they represent two of the most important energy pricing benchmarks and are representative of energy price risks which face energy hedgers. The period examined runs from November 1993 to November 2009. This period was chosen as it contains both tranquil and volatile periods, and is of sufficient length to examine weekly and monthly timeframes.

[9] See http://www.stoxx.com/indices/icb.html for further details.
[10] Contract details are available from http://www.cmegroup.com/trading/energy/crude-oil/light-sweet-crude_contract_specifications.html
[11] Contract details are available at http://www.cmegroup.com/trading/energy/natural-gas/natural-gas_contract_specifications.html

Cash and futures closing prices were obtained from Datastream. Two different frequencies were examined; 5-day (weekly) and 20-day (monthly). We examine these two frequencies as they allow us to compare hedgers using two time horizons that reflect typical investor holding period's while also allowing for sufficient data to carry out a robust analysis. In each case, the returns were calculated as the differenced logarithmic prices over the respective frequencies. Descriptive statistics for the data are displayed in Table 1. Figures 1a and 1b plot the return and volatility patterns of the two series.

[TABLE 1 HERE]

[FIGURE 1A HERE]

[FIGURE 1B HERE]

The following properties of the data are worth noting. Examining first the natural gas series, for both weekly and monthly frequencies there is a positive mean, reflecting the upwards price trend for energy for this period. We can also see that the total risk as measured by standard deviation is quite high, at 10.4% for the weekly spot and 16.8% for the monthly spot returns. From Fig.1a we can see that the size of the volatility has been influenced by a number of spikes, most notably during January 1996. These spikes in the price and associated volatility relate to supply concerns and demand increases associated with severe weather conditions, and the inelastic nature of the natural gas market[12]. Also worth noting is the difference distributional characteristics of the data at the difference frequencies, in particular, monthly data is normally distributed whereas the weekly data is significantly non-normal. This is of relevant to utility based hedgers in that where data is normal, hedging strategies may be broadly similar across different utility functions whereas significant skewness and excess kurtosis will affect the hedging strategies and cause them to diverge for different utility functions such as the quadratic and exponential. Also of note is that the weekly series displays significant positive skewness and excess kurtosis which is associated with extreme movements in the price of natural gas at high frequencies over the period. These findings are standard in describing energy data. Also these distributional properties are of importance when we examine hedging for our economic agents with differing utility functions.

Turning to the Oil contract, that characteristics of the data are broadly similar to Natural Gas. Again we observe a positive mean and significant volatility although it is not as pronounced as for the Natural Gas contract. The most significant volatility was observed during December 2008 following a dramatic fall in the price of Oil, triggered by concerns about the world economy following the financial crisis. There are similarities here between Oil price and Natural Gas price volatility in that both assets demonstrate high degrees of short-term price responsiveness. This means that change to supply or demand conditions can trigger large spot price adjustments. We also note significant kurtosis at both weekly and monthly frequencies. Again of interest is that for the Oil contract, the weekly data is significantly non-normal whereas the monthly data is much closer to normality. In almost all cases there are significant ARCH effects, which support the use of a conditional heteroskedasticity model to estimate the variance covariance matrices of the different energy assets.

[12] Natural gas prices are particularly sensitive to short-term supply and demand shifts for a number of reasons. Significant lead time is required in order to bring additional natural gas supplies to market and to increase pipeline capacity. See Henning, Sloane, and deLeon, (2003).

### 5.1 Estimation

We follow a two stage estimation procedure. We first obtain the risk aversion parameter for energy market participants by fitting a GARCH-M model Eq. (14) to the DJSTOXX Oil and Gas Producers Index data. We then estimate the parameters for the OHR's for Oil and Natural Gas hedgers using spot and futures data, for the period from November 1993 until January 2003. We allow the OHR's to vary over time by the use of a rolling window approach with a window of approximately 10 years to allow for robust estimation of the OHR at both weekly and monthly frequencies. The variance covariance matrix is estimated from the parameters of the DVECH GARCH model. This approach allows us to compare hedges on the basis of utility as distinct from different modelling approaches as the variance covariance matrix underlying the optimal hedges is the same for each utility. After the first OHR is estimated the sample is rolled forward by one observation keeping the window length unchanged. This approach allows us to generate 238 1-period hedges for time t at the weekly frequency and 60 hedges at the monthly frequency. These hedges constitute the in-sample period which stretches from January 2003 until August 2007. We also estimate 1-step ahead forecast hedges for use in period t+1 by using the estimates from the t-period hedges. For this, we reserved a sub-period of data from August 2007 to November 2009 which allows us to estimate 118 out-of-sample hedges at the weekly frequency and 29 at the monthly frequency. The forecast of the risk aversion parameter and the expected return on futures were postulated to follow an AR (1) process (supported in pre-fitting), while the variance and covariance forecasts were derived from the GARCH model parameters.

## 6. Empirical Findings

In this section we examine our findings for two different frequencies, weekly and monthly. We first look at the risk aversion of hedging market participants based on the DJSTOXX Oil and Gas Producers index. We then examine the optimal hedge strategies obtained using those risk aversion parameters for each of the three utility functions, Quadratic, Exponential and Log together with the Minimum Variance Hedging Ratio. Finally we look at hedging effectiveness using both Variance and Value at Risk as our two effectiveness metrics.

### 6.1 Risk Aversion of Energy Hedging Market Participants

Figure 2 and Table 2 present time varying risk aversion parameters for both weekly and monthly hedging frequencies. From Figure 2, we observe a number of distinct features. Firstly the risk aversion parameter is strongly positive for both weekly and monthly frequencies. Risk aversion is also time-varying. Both of these findings have been well documented in the literature on risk aversion for equity markets (see, for example, Brandt and Wang, 2003), but here we confirm them for Oil and Gas market participants. From Figure 2, there is also evidence of a shift in risk aversion. The mean CRRA for the period 2003-2007 of 2.79 is significantly different from the CRRA for the period 2008-2009 of just 2.15 (t-stat 5.58). This shift in risk aversion coincides with the drop in US Industrial Output and Production for 2008 and 2009 and may indicate a pro-cyclical link between the business cycle and the risk aversion of energy market participants. These findings contrast with Brandt and Wang (2003) who find evidence of countercyclical behavior of risk aversion, however, they base their findings on the broader market portfolio as distinct from Oil and Gas producers. Regardless, this result supports the idea that risk aversion for hedging strategies should be based on the observed risk preferences of a particular group of investors rather than taking an average risk aversion for all investors, since they may differ considerably. To the extent that risk aversion drives hedging, then the choice of risk aversion parameter is important.

[FIGURE 2 HERE]

The lower risk aversion for the period 2008 – 2009 indicates that energy investors were prepared to accept a lower expected return for a given level of risk, and may be indicative of a shift in investor and hedger perceptions in 2008 in favor of energy products and away from financial assets in response to the global financial crisis.

In table 2 we examine the summary statistics of the risk aversion parameters. For the weekly frequency we find that risk aversion ranges from 1.73 to 3.44 averaging 2.78. The mean value for the monthly risk aversion parameter at 2.52 is slightly lower, with a range from 0.49 to 3.79. Our results are broadly similar to Ghysels et al (2005) who find the values of the risk aversion parameter in the range 1.5 – 3.3, with an average of 2.7 based on equity market participants. They are also in line with Brandt and Wang (2003) who find a slightly lower average relative risk aversion of about 1.84 based on monthly data in the bond market. Thus, we find that the risk aversion of energy market participants is in line with the broader asset pricing literature. Our findings also indicate that energy market participants display differing attitudes towards risk depending on their investment horizon. Differences in risk aversion between weekly and monthly frequencies are significant at the 1% level. This finding supports Cotter and Hanly (2010) in that different sets of investors may be active at the different frequencies which would explain the differences in observed risk aversion for the different frequencies. It also provides justification for incorporating the risk aversion coefficient into the calculation of the OHR as it allows for the specific risk attitudes of energy investors to dictate the approach to hedging.

[TABLE 2 HERE]

### 6.2 Hedging Strategies

We now examine the Natural Gas and Oil optimal hedging strategies for short and long hedgers for both the weekly and monthly hedging frequencies. Figures 3a (Natural Gas) and 3b (Oil) plot a comparison of the OHR’s for each of the different utility functions,

Quadratic, Log and Exponential together with the Minimum Variance Hedge Ratio. Summary statistics for each of the difference hedge strategies are presented in Table 3. Turning first to the Natural Gas hedges, from Figure 3a we can see that each of the OHR's for both short and long hedger's, displays considerable variation over the time period examined. We can also see large differences in the OHR's for the different utility functions. This indicates the importance of tailoring both the risk aversion parameter and the utility function to the individual investor. From Table 3, we can see the range of the different OHR's. For short hedgers at the weekly frequency, the mean OHR ranges from 0.615 for the Log Utility to 0.803 for the MVHR hedge. For long hedgers the mean OHR ranges from 0.692 for the Exponential Utility to 0.991 for the Log Utility. For hedgers at the monthly frequency the OHR's are generally higher. We also note that in many cases the long hedgers will have an OHR in excess of one. This finding is in line with earlier work by deVille deGoyet, Dhaene and Sercu (2008) and shows that the impact of the expected return together with the risk aversion may result in hedgers increasing their holdings of the futures asset in excess of their cash positions.

[FIGURE 3A HERE]

[FIGURE 3B HERE]

[TABLE 3 HERE]

For the Oil hedgers we again see significant variation in the OHR's for the different utility functions. Looking at both sets of hedgers, for the weekly frequency, the presence of skewness and kurtosis in the data seems to contribute to significant differences between the different utility functions. This is particularly striking for the exponential utility which shows considerable variation when compared to the Quadratic and Log utilities and with the MVHR OHR. This finding is more pronounced for Oil. When we examine the monthly OHR's the Exponential OHR is broadly similar to the Quadratic OHR as we would expect, given that for normal data the utility functions will be the same. These findings support Lien (2007), who notes that while the Quadratic and Exponential Utility functions should yield similar hedges where data is normal, in practice they will produce different optimal hedging decisions as returns data for energy assets is generally characterized by skewness and kurtosis. This finding has important implications for energy hedgers as it again demonstrates that hedgers with different utilities will require different hedging strategies since an approach based on a single utility function will not be optimal.

To further examine the dynamics of the OHR's, we carry out a number of statistical comparisons. Comparing first the weekly with the monthly hedges, we find significant differences for both assets, across all four of the hedge strategies. Their risk aversion may differ, and so too will their optimal hedging strategies. We find that these differences persist irrespective of the utility function thus indicating that investors with different investment horizons will have differing hedging needs in line with their different attitudes towards risk. This further emphasizes that different sets of investors may be active over different holding periods.

Next we examine the Mean OHR for each utility function for short as compared with long hedgers across both frequencies. Taking the weekly frequency for example, for hedgers with quadratic utility, the short hedgers OHR of 0.736 is significantly different than the long hedgers OHR of 0.870. Differences between short and long hedgers are significant for both Natural Gas and Oil, in all cases for the Quadratic, Log and Exponential Utility hedges and across both frequencies. This is not surprising given that each set of hedgers in interested in outcomes from opposite ends of the return distribution. It also emphasizes that incorporating risk aversion into the hedging decision allows the expected return to play a part in the choice of OHR whereas when risk aversion is not explicitly modeled[13], the OHR will be the same for both short and long hedgers.

When we compare the OHR's across the different utility functions a number of interesting results emerge. Table 4, provides a comparison of the absolute differences between the mean of each of the different OHR's. We compare the Mean OHR for each of the different utility functions with each other for each set of hedgers and within each frequency. Taking short hedgers at the weekly frequency, for example, the differences between the Quadratic OHR and the Log, Exponential and MVHR OHR's are 0.12, 0.03 and 0.07 respectively. For Natural Gas, we find significant differences in all cases at the 1% level. For Oil, the differences are significant with the exception of the Exponential Utility at the weekly frequency. If we focus on the difference between the MVHR OHR which assumes infinite risk aversion and the utility based OHR's, we find significant differences in every single case for both Natural Gas and Oil. For example, using Short Oil Hedges at the weekly frequency, the MVHR OHR differs from the Quadratic OHR by 0.13 (t-stat 41.96), from the Log OHR by 0.36 (49.99) and from the Exponential OHR by 0.18 (5.12). These differences are all significant, and we find similar differences for the other frequencies and for the Natural Gas hedges.

[13] When infinite risk aversion is assumed, the utility based hedges converge to the OLS MVHR.

[TABLE 4 HERE]

These findings indicate that the hedging strategy is contingent not just on the dynamics of spot and futures prices but also on the utility function of the hedger. It also indicates that different risk aversion will yield different hedge strategies.  The most similar OHR's tend to be the Quadratic and Exponential OHR's particularly for the monthly frequency. Again, this relates to the fact that monthly returns are more normal than weekly returns. These findings provide further evidence of the importance of incorporating risk aversion into the hedging decision. The results highlight  not just risk aversion but also the utility function and its importance in estimating a hedging strategy, not just for Oil hedgers but for energy hedgers more generally.

### 6.3 Hedging Performance

We turn next to the performance of the different hedges using two different metrics, the Variance and Value at Risk. We first examine the overall hedging performance across all of the difference OHR's in-sample. From Table 5a, for weekly Natural Gas, overall hedging effectiveness using the Variance risk measure averages about 46% across all hedges for the short hedgers and 47% for the long hedgers. For the monthly hedges this increases to 78% and 81% for short and long hedgers respectively.

[TABLE 5a and 5b HERE]

From Table 5b, for Oil hedgers at the weekly frequency, the variance reduction is about 69% for short hedgers and 65% for long hedgers. For the monthly hedges this increases to 88% and 85% for short and long hedgers respectively. The VaR metric is broadly consistent with the Variance in terms of the relative performance of the different hedging strategies. More specifically, reductions in the VaR are of the order of 27% (Weekly) and 55% (Monthly) for Natural Gas and 48% (Weekly) and 70% (Monthly) for Oil.

Comparing next the relative performance of the different utility functions, significant differences emerge in risk reduction. If we look at a $1,000,000 exposure for a long Natural Gas hedger at the monthly frequency for example, the OHR using exponential utility will reduce the VaR to $139,000 whereas for the log utility the VaR reduction is €176,600, a difference of $37,700. Similar differences are found for Oil hedgers and for both weekly and monthly frequencies. This finding shows that there are significant economic differences between hedgers with different utility functions and is indicative of the importance of incorporating utility into the hedging decision.

These findings also show that utility based hedges are effective at reducing risk from an economic perspective, when measured using convention risk metrics such as variance and VaR. In terms of a comparison, performance is markedly better at lower frequencies for both Natural Gas and Oil for all utilities with the possible exception of the log utility. This relates to the higher correlation between spot and futures for monthly data. There is little difference in performance between short and long hedgers. For Natural Gas, long hedgers do marginally better on average in economic terms, whereas for Oil the position is reversed. In terms of the best hedging model, the clear winner is the MVHR. This shows average reductions in variance of the order of 71% across all frequencies for both assets. This is followed by the Quadratic Utility model at 68%. Results are similar for the VaR risk measure. These results relate to the use of the Variance and VaR as performance metrics as they focus on risk alone. The performance of the Log and Exponential Utility models in terms of the Variance and the VaR is still acceptable in economic terms and in one case the Exponential model is the best performer for long Natural Gas hedgers at the monthly frequency.

In terms of the out-of-sample performance, from Tables 5a and 5b, the results are broadly similar to the in-sample results. For weekly hedges the reductions in Variance for both Short and Long hedgers are of the order of 50% for Natural Gas and 85% for Oil. Also VaR reductions are about 30% and 64% for Natural Gas and Oil respectively. For monthly hedges, Natural Gas again shows performance improvements over weekly hedges however, for Oil somewhat surprisingly the hedging performance as measured by Variance and VaR disimproves by about 20% when comparing weekly with monthly hedges. Finally examining the performance of the different models, again the MVHR and Quadratic models are consistently the best performers.

## 7. Conclusion

We estimate and compare utility based optimal hedge strategies based on the risk preferences of energy market participants. By addressing the differing risk attitudes of energy hedgers, we are highlighting an issue that is of real relevance to investors in energy markets at a time when energy price movements are increasingly uncertain and attitudes towards risk have shown dramatic shifts in response to the global financial crisis. We use an approach that allows us to incorporate time varying risk aversion and apply it to time varying hedge ratios that are optimized for a variety of differing utility functions. We apply our approach to both the Crude Oil and Natural Gas markets

Significant differences emerge between the hedge strategies depending on the risk attitudes of energy hedgers as represented by different utility functions. These differences are particularly pronounced for non-normal data as characterized by skewness and kurtosis. Since this tends to describe energy assets such as Oil and Natural Gas, the implication for hedgers is that they should optimise their hedges by explicitly taking their own risk preferences and utility into account. Our results also indicate that energy market participants exhibit risk aversion that is broadly similar to that found in the asset pricing literature and in particular to the equity market. Furthermore, the risk attitudes of investors tend to vary over time and this is particularly true for the recent timeframe.

Finally we note that these changes in attitudes towards risk are of particular relevance to energy hedgers and further work in this area could yield fresh insights into ways of addressing the hedging needs of energy market participants.

## References

Alexander, C., 2008. Quantitative Methods in Finance, Wiley, England.

Arrow, K., 1971. Essays in the theory of risk bearing, Markham, New York.

Bernoulli D 1954 Exposition of a new theory on the measurement of risk; Econometrica 22 23–36 (Translation of Bernoulli D 1738 Specimen theoriae novae de mensura sortis; Papers Imp. Acad. Sci. St. Petersburg 5 175–192)

Bollerslev, T., Engle, R., Wooldridge, J., 1988. A capital asset pricing model with time-varying covariances. Journal of Political Economy 96, 116 - 131.

Brandt, M., Wang, K., 2003. Time-varying risk aversion and unexpected inflation. Journal of Monetary Economics 50, 1457-1498.

Brooks, C., Cerny, A., Miffre, J., 2007. Optimal hedging with higher moments. Working Paper, Cass Business School, City University London.

Cabedo, J., Moya, I., 2003. Estimating oil price Value at Risk using the historical simulation approach. Energy Economics 25, 239-253.

Cecchetti, S.G., Cumby, R.E., & Figlewski, S. 1988. Estimation of the optimal futures hedge. Review of Economics and Statistics 70, 623 – 630.

Chen, K., Sears, R., Tzang., D 1987. Oil prices and energy futures. Journal of Futures Markets, 7, 501 – 518.

Cotter, J., Hanly, J., 2006. Re-examining hedging performance. Journal of Futures Markets 26, 657-676.

Cotter, J., Hanly, J., 2010. Time-varying risk aversion: an application to energy hedging. Energy Economics, 32, 432 - 441.

Danthine, J and Donaldson, J., 2005, Intermediate financial Theory. Elsevier

deVille deGoyet, C., Dhaene, G., Sercu, P., 2008. Testing the martingale hypothesis for futures prices: Implications for hedgers. Journal of Futures Markets 28, 1040 – 1065.

Ederington, L. (1979). The hedging performance of the new futures markets. Journal of Finance 34, 157 – 170.

Engle, R., Lilien, D., Robbins, R., 1987. Estimating time varying risk premia in the term structure: the arch-m model. Econometrica 55, 391 – 407.

Ghysels, E., Santa-Clara, P., Valkanov, R., 2005. There is a risk return trade-off after all. Journal of Financial Economics 76, 509 - 548.

Giovannini, A., Jorion, P., 1989. The time variation of risk and return in the foreign exchange and stock markets. Journal of Finance 2, 307 – 325.

Henning, B., Sloane, M., deLeon, M., 2003. Natural Gas and Energy Price Volatility, Arlington, Virginia, Energy and Environmental Analysis Inc.

Kroner, K.F., & Sultan, J. 1993. Time varying distribution and dynamic hedging with foreign currency futures. Journal of Financial and Quantitative Analysis, 28, 535-551.

Lien, D., 2007. Optimal futures hedging: quadratic versus exponential utility functions. Journal of Futures Markets 28, 208 - 211.

Markowitz, H., 1952, Portfolio Selection, Journal of Finance 7, 77-91.

Mastrangelo, D., 2007. An Analysis of Price Volatility in Natural Gas Markets. U.S. Energy Information Administration. Office of Oil and Gas.

Merton, R., 1980. On estimating the expected return on the market. Journal of Financial Economics 8, 323–361.

Merton, R., 1990. Continuous Time Finance. Blackwell. Cambridge. MA.

Regnier, E., 2007. Oil and energy price volatility. Energy Economics 29, 405 – 427.

Townsend, R. 1994. Risk and Insurance in Village India. Econometrica, 62, 539-591.

| Index | Frequency | Mean | Min | Max | Std Dev | Skewness | Kurtosis | B-J | LM | Stationarity |
|---|---|---|---|---|---|---|---|---|---|---|
| **NATURAL GAS** | | | | | | | | | | |
| Spot | 5-day | 0.0005 | -0.73 | 0.94 | 0.104 | 0.499* | 13.14* | 6051.2* | 203.7* | -14.28* |
| Futures | 5-day | 0.0007 | -0.37 | 0.40 | 0.082 | 0.244* | 2.09* | 160.2* | 37.6* | -13.15* |
| Spot | 20-day | 0.0022 | -0.55 | 0.46 | 0.168 | -0.183 | 0.73 | 5.81 | 11.9** | -6.80* |
| Futures | 20-day | 0.0029 | -0.46 | 0.49 | 0.153 | -0.031 | 0.50 | 2.22 | 6.19 | -6.77* |
| **OIL** | | | | | | | | | | |
| Spot | 5-day | 0.0018 | -0.23 | 0.30 | 0.054 | -0.117 | 2.15* | 163.4* | 56.8* | -11.83* |
| Futures | 5-day | 0.0018 | -0.24 | 0.23 | 0.049 | -0.151 | 1.70* | 103.5* | 49.1* | -11.49* |
| Spot | 20-day | 0.0072 | -0.33 | 0.27 | 0.098 | -0.690* | 1.23* | 26.0* | 11.2** | -6.53* |
| Futures | 20-day | 0.0072 | -0.32 | 0.27 | 0.095 | -0.613* | 0.91* | 19.3* | 11.2** | -6.42* |
| 1% C.V | | | | | | | | 9.21 | 13.23 | -3.44 |

**Table 1: Summary Statistics of Natural Gas and Oil**

Summary statistics are presented for the log returns of each spot and futures series. The skewness statistic measures asymmetry where zero would indicate a symmetric distribution. The total sample period runs from 03/11/1993 until 04/11/2009.The kurtosis statistic measures the shape of a distribution as compared with a normal distribution. Figures reported for kurtosis are for excess kurtosis where a value of zero would indicate a normal distribution. The Bera-Jarque (B-J) statistic combines skewness and kurtosis to measure normality. LM, (with 4 lags) is the Lagrange Multiplier test proposed by Engle (1982). The test statistics for B-J and LM tests are distributed $\chi 2$. Stationarity is tested using Augmented the Dickey-Fuller unit root test with 4 lags. This is important as it ensures that the relationship between the spot and futures assets is robust. *Denotes Significance at the 1% level. **Denotes Significance at the 5% level.

| | CRRA | |
|---|---|---|
| | **WEEKLY** | **MONTHLY** |
| MEAN | 2.78† | 2.52 |
| MIN | 1.73 | 0.49 |
| MAX | 3.44 | 3.79 |
| STDEV | 0.37 | 0.48 |

**Table 2: Risk Aversion of Short and Long Hedgers**

CRRA is the estimated risk aversion parameter, summary statistics are presented for the in sample period. Statistical comparisons are drawn between the Mean CRRA value for the weekly and monthly hedging intervals. There are significant differences between the CRRA values at weekly and monthly frequencies. † denotes significance at the 1% level for comparison of the CRRA for weekly Vs monthly frequencies.

| | PANEL A: SHORT HEDGERS | | | | PANEL B: LONG HEDGERS | | | |
|---|---|---|---|---|---|---|---|---|
| | (1) | (2) | (3) | (4) | (5) | (6) | (7) | (8) |
| | OHR - QUAD | OHR - LOG | OHR - EXP | OHR - MVHR | OHR - QUAD | OHR - LOG | OHR - EXP | OHR - MVHR |
| | | | | NATURAL GAS | | | | |
| **WEEKLY** | | | | | | | | |
| MEAN | 0.736†* | 0.615†* | 0.705†* | 0.803† | 0.870† | 0.991† | 0.692† | 0.803† |
| MIN | 0.050 | 0.030 | 0.239 | 0.065 | 0.080 | 0.100 | 0.248 | 0.065 |
| MAX | 0.981 | 0.859 | 1.129 | 1.052 | 1.123 | 1.326 | 1.392 | 1.052 |
| STDEV | 0.137 | 0.127 | 0.166 | 0.146 | 0.157 | 0.183 | 0.266 | 0.146 |
| **MONTHLY** | | | | | | | | |
| MEAN | 0.904* | 0.700* | 0.887* | 1.010 | 1.117 | 1.320 | 1.096 | 1.010 |
| MIN | 0.759 | 0.444 | 0.799 | 0.838 | 0.917 | 1.050 | 1.073 | 0.838 |
| MAX | 1.132 | 0.963 | 0.956 | 1.234 | 1.336 | 1.568 | 1.132 | 1.234 |
| STDEV | 0.063 | 0.093 | 0.031 | 0.059 | 0.065 | 0.099 | 0.015 | 0.059 |
| | | | | OIL | | | | |
| **WEEKLY** | | | | | | | | |
| MEAN | 0.878†* | 0.648†* | 0.824†* | 1.008† | 1.138† | 1.368† | 1.182† | 1.008† |
| MIN | 0.694 | 0.295 | -0.921 | 0.805 | 0.916 | 1.016 | 0.142 | 0.805 |
| MAX | 0.997 | 0.924 | 1.893 | 1.078 | 1.274 | 1.726 | 2.930 | 1.078 |
| STDEV | 0.043 | 0.109 | 0.557 | 0.020 | 0.045 | 0.114 | 0.560 | 0.020 |
| **MONTHLY** | | | | | | | | |
| MEAN | 0.813* | 0.453* | 0.706* | 1.013 | 1.213 | 1.572 | 1.337 | 1.013 |
| MIN | 0.621 | 0.056 | 0.320 | 0.959 | 1.065 | 1.192 | 1.128 | 0.959 |
| MAX | 0.932 | 0.805 | 0.902 | 1.050 | 1.406 | 1.971 | 1.729 | 1.050 |
| STDEV | 0.081 | 0.165 | 0.129 | 0.013 | 0.083 | 0.169 | 0.135 | 0.013 |

**Table 3: Optimal Hedge Strategies of Short and Long Hedgers**

Summary statistics are presented for the in-sample period for both Natural Gas and Oil at weekly and monthly hedging intervals for both short and long hedgers. Two statistical comparisons are drawn. We first compare the mean OHR's for weekly and monthly intervals. Using Natural Gas for example, we find a significant difference between the Quadratic OHR (column 1) for a weekly hedger (0.736) with that of a monthly hedger (0.904). We also compare the mean hedge ratios of short Vs long hedgers. Using the Oil hedges at the weekly frequency for example, we find a significant difference between the Log OHR (0.648) for a short hedger and the Log OHR (1.368) for a long hedger. * denotes significance at the 1% level respectively for short Vs long comparison. † denotes significance at the 1% level for comparison of weekly and monthly OHR's.

| | PANEL A: SHORT HEDGERS | | | | PANEL B: LONG HEDGERS | | | |
|---|---|---|---|---|---|---|---|---|
| | | | | **NATURAL GAS** | | | | |
| | **(1)** | **(2)** | **(3)** | **(4)** | **(5)** | **(6)** | **(7)** | **(8)** |
| | **QUAD** | **LOG** | **EXP** | **MVHR** | **QUAD** | **LOG** | **EXP** | **MVHR** |
| WEEKLY | | | | | | | | |
| OHR - QUAD | 0.00 | 0.12* | **0.03*** | 0.07* | 0.00 | 0.12* | 0.18* | **0.07*** |
| | | (9.99) | (2.22) | (5.19) | | (7.76) | (8.92) | (4.85) |
| OHR - LOG | | 0.00 | 0.09* | 0.19* | | 0.00 | 0.30* | 0.19* |
| | | | (6.65) | (15.01) | | | (14.30) | (12.42) |
| OHR - EXP | | | 0.00 | 0.10* | | | 0.00 | 0.11* |
| | | | | (6.87) | | | | (5.65) |
| OHR - MVHR | | | | 0.00 | | | | 0.00 |
| | | | | | | | | |
| MONTHLY | | | | | | | | |
| OHR - QUAD | 0.00 | 0.20* | **0.02*** | 0.11* | 0.00 | 0.20* | **0.02*** | 0.11* |
| | | (27.56) | (3.73) | (18.66) | | (26.11) | (4.59) | (18.29) |
| OHR - LOG | | 0.00 | 0.19* | 0.31* | | 0.00 | 0.22* | 0.31* |
| | | | (28.88) | (42.70) | | | (33.99) | (40.82) |
| OHR - EXP | | | 0.00 | 0.12* | | | 0.00 | 0.09* |
| | | | | (28.02) | | | | (21.37) |
| OHR - MVHR | | | | 0.00 | | | | 0.00 |
| | | | | **OIL** | | | | |
| | **QUAD** | **LOG** | **EXP** | **MVHR** | **QUAD** | **LOG** | **EXP** | **MVHR** |
| WEEKLY | | | | | | | | |
| OHR - QUAD | 0.00 | 0.23* | **0.05** | 0.13* | 0.00 | 0.23* | **0.04** | 0.13* |
| | | (30.25) | (1.52) | (41.96) | | (28.87) | (1.20) | (40.33) |
| OHR - LOG | | 0.00 | 0.18* | 0.36* | | 0.00 | 0.19* | 0.36* |
| | | | (4.79) | (49.99) | | | (5.05) | (47.79) |
| OHR - EXP | | | 0.00 | 0.18* | | | 0.00 | 0.17* |
| | | | | (5.12) | | | | (4.79) |
| OHR - MVHR | | | | 0.00 | | | | 0.00 |
| | | | | | | | | |
| MONTHLY | | | | | | | | |
| OHR - QUAD | 0.00 | 0.36* | **0.11*** | 0.20* | 0.00 | 0.36* | **0.12*** | 0.20* |
| | | (30.29) | (11.92) | (38.59) | | (29.77) | (13.42) | (37.82) |
| OHR - LOG | | 0.00 | 0.25* | 0.56* | | 0.00 | 0.23* | 0.56* |
| | | | (19.52) | (52.05) | | | (17.61) | (51.20) |
| OHR - EXP | | | 0.00 | 0.31* | | | 0.00 | 0.32* |
| | | | | (41.61) | | | | (41.93) |
| OHR - MVHR | | | | 0.00 | | | | 0.00 |

**Table 4: Comparison of Differences between Optimal Hedge Strategies for Different Utilitiy Functions**

Table 4 presents a comparison of the differences between the Mean OHR's of the different utility functions. Using short hedgers at the weekly frequency for Oil as an example, the difference between the Quadratic and the Exponential OHR's is 0.05. This difference is not significant (t-statistics are in parentheses). The most similar hedges defined as those with the smallest difference in the Mean OHR are highlighted in black. * denotes significance at the 1% level for comparison of Mean OHR's between different utility functions. The quadratic utility function is used as a benchmark given its wide use in asset pricing and portfolio applications.

| | Panel A: Short Hedgers | | | | | Panel B: Long Hedgers | | | | |
|---|---|---|---|---|---|---|---|---|---|---|
| | (1) HE ($x10^{-2}$) | (2)HE ($x10^{-2}$) | (3)HE ($x10^{-2}$) | (4)HE ($x10^{-2}$) | (5)HE ($x10^{-2}$) | (6)HE ($x10^{-2}$) | (7)HE ($x10^{-2}$) | (8)HE ($x10^{-2}$) | (9)HE ($x10^{-2}$) | (10)HE ($x10^{-2}$) |
| | **OHR - QUAD** | **OHR - LOG** | **OHR - EXP** | **OHR - OLS** | **NO HEDGE** | **OHR - QUAD** | **OHR - LOG** | **OHR - EXP** | **OHR - OLS** | **NO HEDGE** |
| **IN-SAMPLE** | | | | | | | | | | |
| **WEEKLY** | | | | | | | | | | |
| MEAN | -0.0450 | -0.0384 | 0.1389 | -0.0557 | 0.0741 | 0.0664 | 0.0731 | -0.0584 | 0.0557 | -0.0741 |
| VARIANCE | 0.5288 | 0.5654 | 0.5757 | 0.5128 | 1.0175 | 0.5034 | 0.5065 | 0.6252 | 0.5128 | 1.0175 |
| VaR 1% | -16.96 | -17.53 | -17.51 | -16.71 | -23.39 | -16.44 | -16.48 | -18.45 | -16.60 | -23.54 |
| HE1 | 48.03 | 44.43 | 43.42 | 49.60 | 0.00 | 50.52 | 50.22 | 38.55 | 49.60 | 0.00 |
| HE2 | 27.49 | 25.06 | 25.13 | 28.55 | 0.00 | 30.16 | 29.98 | 21.61 | 29.47 | 0.00 |
| **MONTHLY** | | | | | | | | | | |
| MEAN | -0.0387 | -0.1580 | 0.0390 | 0.0405 | 0.0448 | -0.1197 | -0.2390 | -0.1084 | -0.0405 | -0.0448 |
| VARIANCE | 0.4258 | 0.7352 | 0.4377 | 0.3632 | 2.2665 | 0.3689 | 0.5614 | 0.3515 | 0.3632 | 2.2665 |
| VaR 1% | -15.22 | -20.10 | -15.35 | -13.98 | -34.98 | -14.25 | -17.67 | -13.90 | -14.06 | -35.07 |
| HE1 | 81.22 | 67.56 | 80.69 | 83.97 | 0.00 | 83.72 | 75.23 | 84.49 | 83.97 | 0.00 |
| HE2 | 56.49 | 42.52 | 56.11 | 60.03 | 0.00 | 59.37 | 49.61 | 60.36 | 59.90 | 0.00 |
| **OUT-OF-SAMPLE** | | | | | | | | | | |
| **WEEKLY** | | | | | | | | | | |
| MEAN | -0.3017 | -0.3082 | -0.1529 | -0.2450 | -0.4803 | 0.1882 | 0.1817 | 0.5262 | 0.2450 | 0.4803 |
| VARIANCE | 0.2996 | 0.3039 | 0.3319 | 0.3058 | 0.6583 | 0.3164 | 0.3318 | 0.4139 | 0.3058 | 0.6583 |
| VaR 1% | -13.04 | -13.13 | -13.56 | -13.11 | -19.36 | -12.90 | -13.22 | -14.44 | -12.62 | -18.39 |
| HE1 | 54.48 | 53.83 | 49.58 | 53.55 | 0.00 | 51.94 | 49.60 | 37.13 | 53.55 | 0.00 |
| HE2 | 32.65 | 32.15 | 29.96 | 32.27 | 0.00 | 29.89 | 28.14 | 21.50 | 31.40 | 0.00 |
| **MONTHLY** | | | | | | | | | | |
| MEAN | -1.2673 | -1.2172 | -1.8464 | -0.8846 | -0.7382 | 0.5018 | 0.5519 | 0.2926 | 0.8846 | 0.7382 |
| VARIANCE | 0.1950 | 0.2168 | 0.2507 | 0.2164 | 1.7101 | 0.2771 | 0.3321 | 0.4484 | 0.2164 | 1.7101 |
| VaR 1% | -11.54 | -12.05 | -13.50 | -11.71 | -31.16 | -11.74 | -12.85 | -15.29 | -9.94 | -29.68 |
| HE1 | 88.59 | 87.32 | 85.34 | 87.35 | 0.00 | 83.79 | 80.58 | 73.78 | 87.35 | 0.00 |
| HE2 | 62.96 | 61.33 | 56.69 | 62.43 | 0.00 | 60.43 | 56.69 | 48.50 | 66.52 | 0.00 |

**Table 5a: Hedged Returns and Hedging Performance – Natural Gas**

Mean, Variance, VaR and Hedging Effectiveness (HE) are presented for the each of the hedging strategies, HE1 is a measure of the percentage reduction in the Variance from each of the hedging methods as compared with a No hedge position (or the worst performing hedge strategy). For example, for a weekly short hedger in-sample, the OLS OHR reduces the variance by 49.60% as compared with a no hedge position. Similarly, HE2 measures the percentage reduction in the 1% VaR.

| | Panel A: Short Hedgers | | | | | Panel B: Long Hedgers | | | | |
|---|---|---|---|---|---|---|---|---|---|---|
| | (1) HE $(x10^{-2})$ | (2)HE $(x10^{-2})$ | (3)HE $(x10^{-2})$ | (4)HE $(x10^{-2})$ | (5)HE $(x10^{-2})$ | (6)HE $(x10^{-2})$ | (7)HE $(x10^{-2})$ | (8)HE $(x10^{-2})$ | (9)HE $(x10^{-2})$ | (10)HE $(x10^{-2})$ |
| | **OHR - QUAD** | **OHR - LOG** | **OHR - EXP** | **OHR - OLS** | **NO HEDGE** | **OHR - QUAD** | **OHR - LOG** | **OHR - EXP** | **OHR - OLS** | **NO HEDGE** |
| **IN-SAMPLE** | | | | | | | | | | |
| **WEEKLY** | | | | | | | | | | |
| MEAN | 0.0215 | 0.0799 | -0.0945 | 0.0023 | 0.3262 | 0.0169 | 0.0753 | -0.1011 | -0.0023 | -0.3262 |
| VARIANCE | 0.0259 | 0.0449 | 0.1519 | 0.0234 | 0.1996 | 0.0284 | 0.0497 | 0.1702 | 0.0234 | 0.1996 |
| VaR 1% | -3.72 | -4.85 | -9.16 | -3.56 | -10.07 | -3.90 | -5.11 | -9.70 | -3.56 | -10.72 |
| HE1 | 87.02 | 77.49 | 23.89 | 88.27 | 0.00 | 85.78 | 75.10 | 14.72 | 88.27 | 0.00 |
| HE2 | 63.01 | 51.81 | 8.99 | 64.66 | 0.00 | 63.60 | 52.32 | 9.52 | 66.77 | 0.00 |
| **MONTHLY** | | | | | | | | | | |
| MEAN | 0.0654 | 0.3570 | 0.1034 | -0.0291 | 1.2250 | 0.1235 | 0.4152 | 0.1463 | 0.0291 | -1.2250 |
| VARIANCE | 0.0300 | 0.2328 | 0.0757 | 0.0020 | 0.7121 | 0.0423 | 0.2663 | 0.1022 | 0.0020 | 0.7121 |
| VaR 1% | -3.96 | -10.87 | -6.30 | -1.07 | -18.41 | -4.66 | -11.59 | -7.29 | -1.01 | -20.86 |
| HE1 | 95.79 | 67.31 | 89.37 | 99.72 | 0.00 | 94.06 | 62.61 | 85.64 | 99.72 | 0.00 |
| HE2 | 78.46 | 40.96 | 65.79 | 94.19 | 0.00 | 77.66 | 44.43 | 65.04 | 95.15 | 0.00 |
| **OUT-OF-SAMPLE** | | | | | | | | | | |
| **WEEKLY** | | | | | | | | | | |
| MEAN | -0.015 | -0.049 | 0.325 | -0.009 | 0.078 | 0.004 | -0.030 | 0.359 | 0.009 | -0.078 |
| VARIANCE | 0.045 | 0.164 | 0.081 | 0.020 | 0.541 | 0.043 | 0.155 | 0.115 | 0.020 | 0.541 |
| VaR 1% | -4.94 | -9.48 | -6.29 | -3.29 | -17.04 | -4.80 | -9.18 | -7.52 | -3.27 | -17.19 |
| HE1 | 91.72 | 69.61 | 85.07 | 96.33 | 0.00 | 92.13 | 71.41 | 78.79 | 96.33 | 0.00 |
| HE2 | 71.00 | 44.34 | 63.09 | 80.69 | 0.00 | 72.10 | 46.60 | 56.24 | 80.98 | 0.00 |
| **MONTHLY** | | | | | | | | | | |
| MEAN | -0.70 | -0.34 | -2.37 | -0.03 | 0.47 | -0.64 | -0.29 | -2.33 | 0.03 | -0.47 |
| VARIANCE | 0.34 | 0.70 | 2.10 | 0.00 | 1.64 | 0.32 | 0.70 | 2.06 | 0.00 | 1.64 |
| VaR 1% | -14.17 | -19.82 | -36.07 | -1.53 | -29.30 | -13.86 | -19.79 | -35.75 | -1.47 | -30.23 |
| HE1 | 84.01 | 66.62 | 0.00 | 99.80 | 21.99 | 84.35 | 65.95 | 0.00 | 99.80 | 20.67 |
| HE2 | 60.71 | 45.07 | 0.00 | 95.77 | 18.77 | 61.22 | 44.63 | 0.00 | 95.89 | 15.42 |

**Table 5b: Hedged Returns and Hedging Performance – Oil**

Mean, Variance, VaR and Hedging Effectiveness (HE) are presented for the each of the hedging strategies, HE1 is a measure of the percentage reduction in the Variance from each of the hedging methods as compared with a No hedge position (or the worst performing hedge strategy). For example, for a weekly short hedger in-sample, the OLS OHR reduces the variance by 88.27% as compared with a no hedge position. Similarly, HE2 measures the percentage reduction in the 1% VaR.

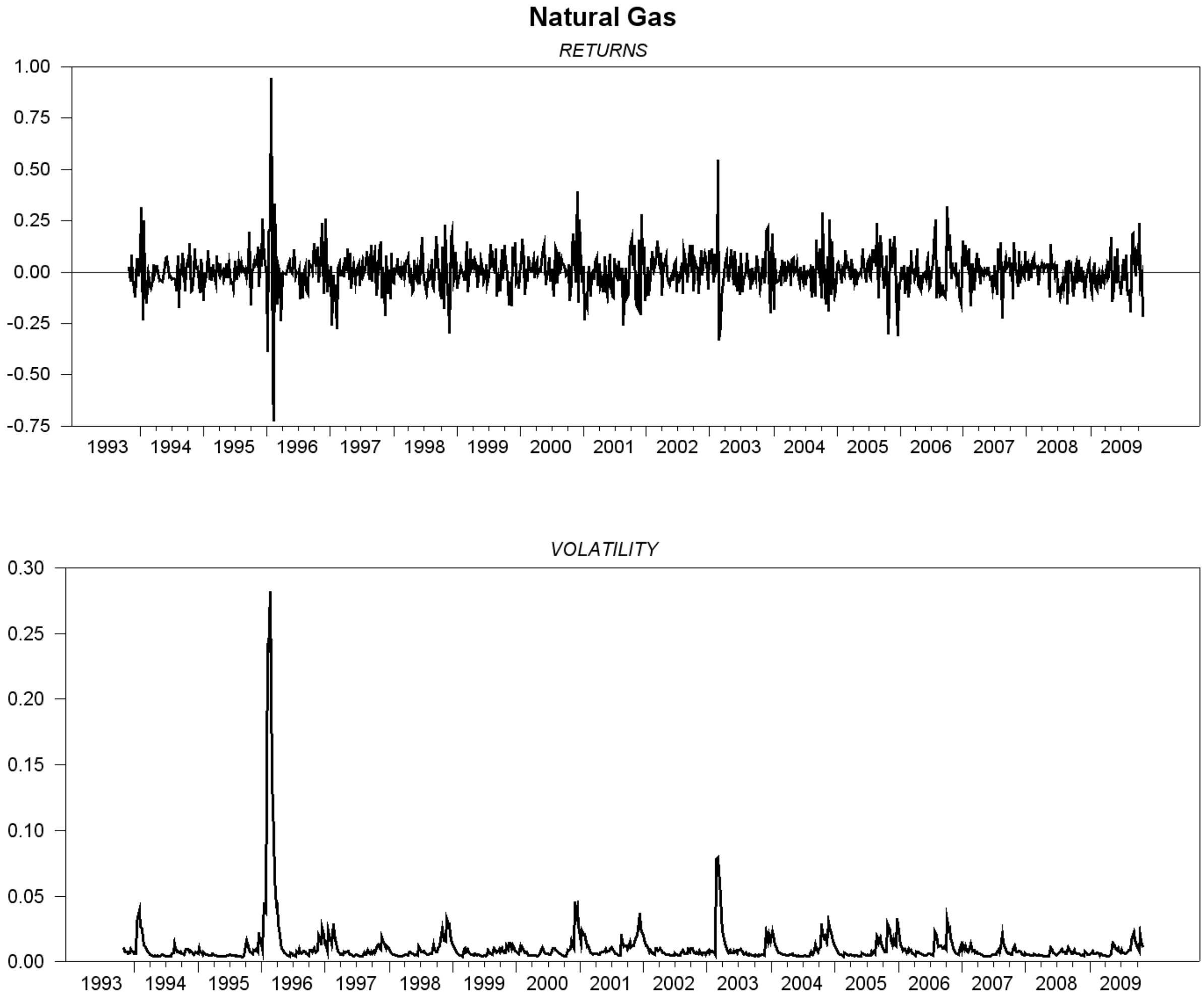

**Fig 1a:Time Series Plot**

Fig 1a displays the general data characteristics for the weekly natural gas spot series. Each series is shown for the period from 03/11/1993 to 04/11/2009. Volatility is obtained from fitting a GARCH (1, 1) model. The large spike in volatility in January 1996 is attributed to supply concerns as inventories dropped to 35% b below the 5-year average[14].

[14] See analysis carried out by the Energy Information Administration, 2007.

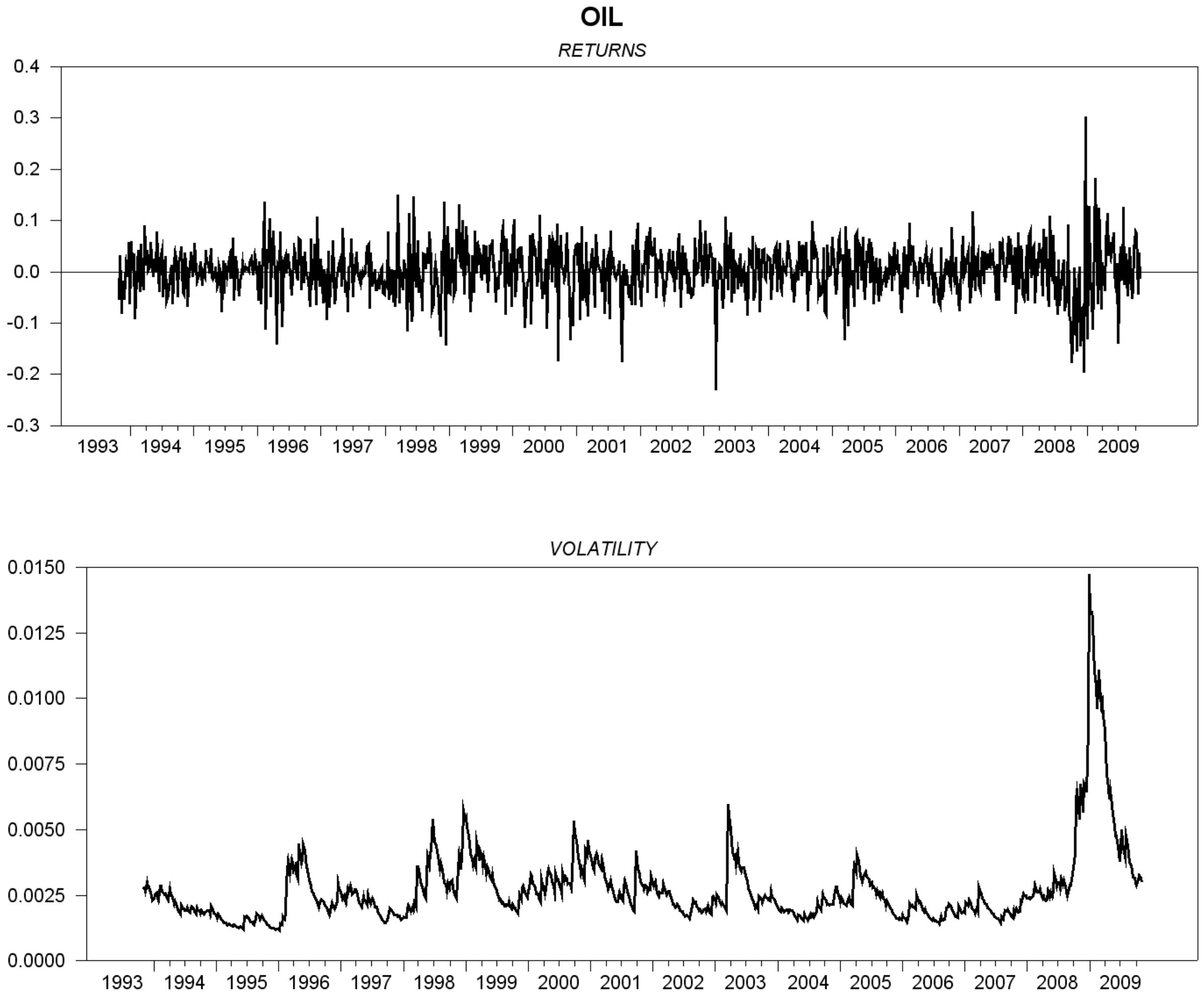

**Fig 1b: Time Series Plot**

Fig 1a displays the general data characteristics for the weekly natural gas spot series. Each series is shown for the period from 03/11/1993 to 04/11/2009. Volatility is obtained from fitting a GARCH (1, 1) model. The large spike in volatility in January 1996 is attributed to supply concerns as inventories dropped to 35% b below the 5-year average[15].

[15] See analysis carried out by the Energy Information Administration, 2007.

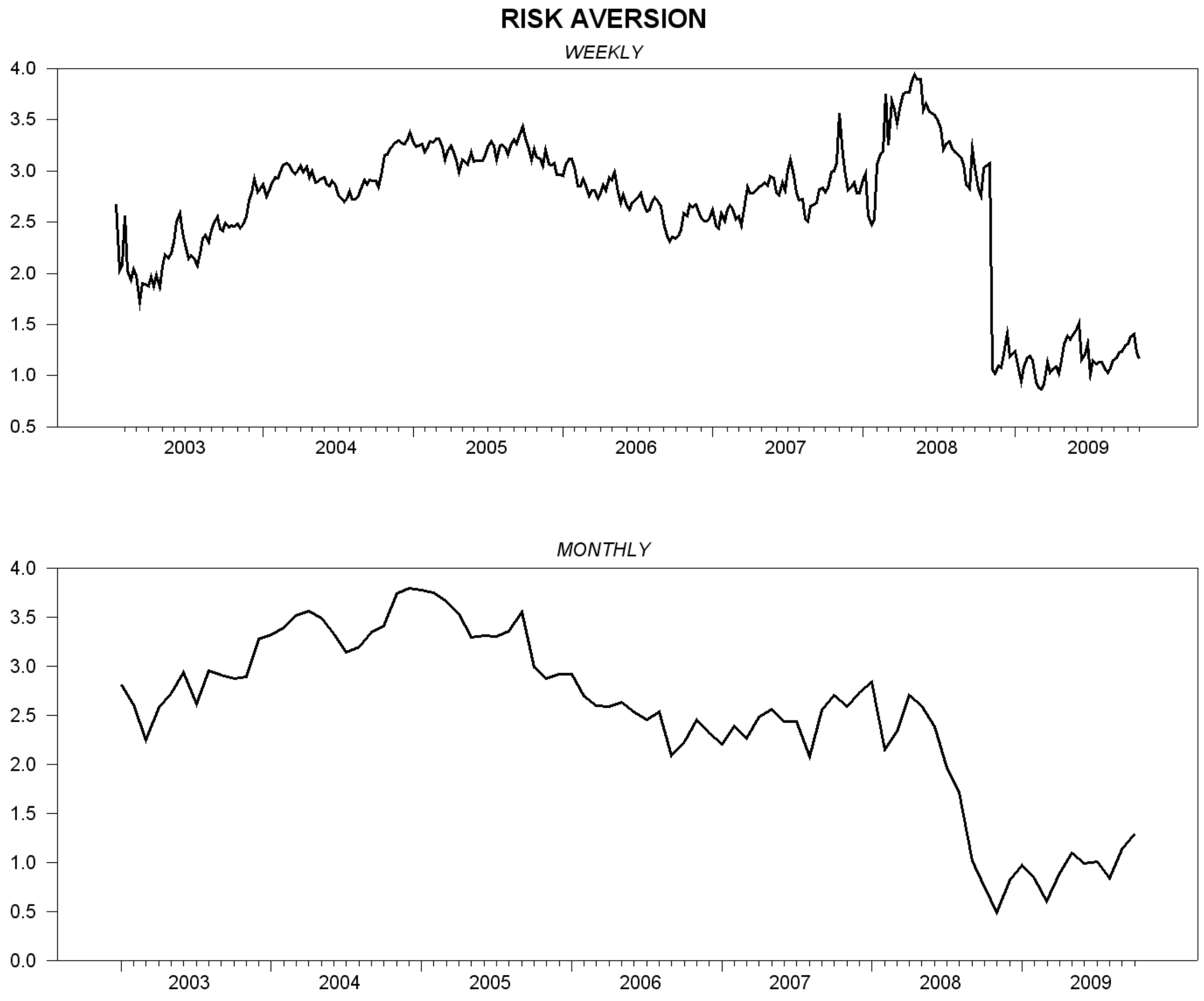

**Fig 2: Time-varying Coefficient of Relative Risk Aversion**

The CRRA is plotted for the weekly and monthly hedging intervals. The risk aversion is based on the risk and return characteristics of the Oil and Gas Producers Index.

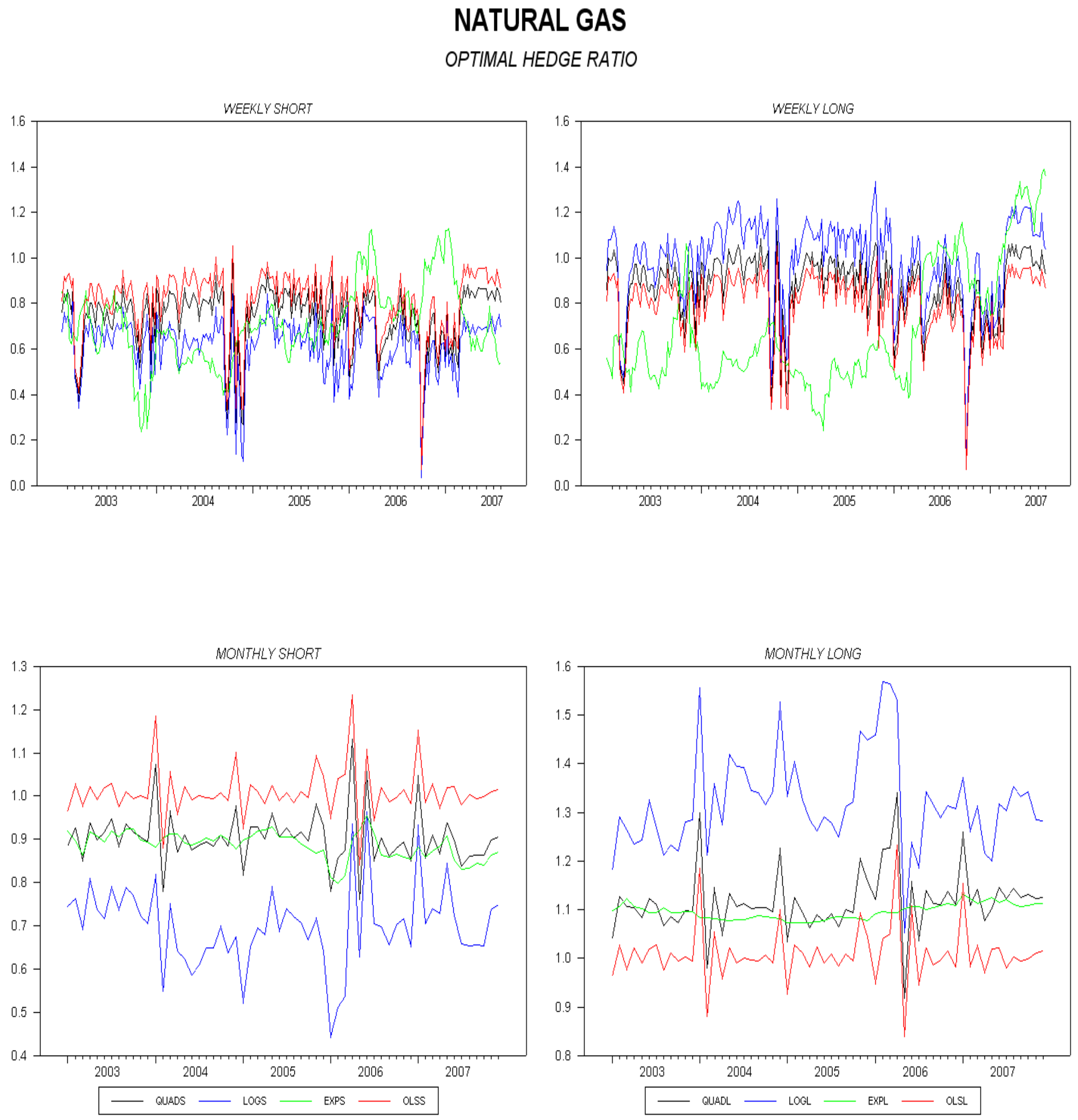

**Fig 3a:Time-varying Optimal Hedge Ratios for the Natural Gas contract**

This figure is a time-series plot of the OHR's for the in-sample period for each of the different utility functions, Quadratic, Log, and Exponential together with the MVHR OHR for both short and long hedgers for both weekly and monthly hedging frequencies. The variance covariance matrix is calculated using a rolling window to allow each of the hedge strategies to vary over time. This allows us to compare hedges on the basis of utility as the variance covariance matrix underlying the optimal hedges is the same for each utility.

**OIL**

*OPTIMAL HEDGE RATIO*

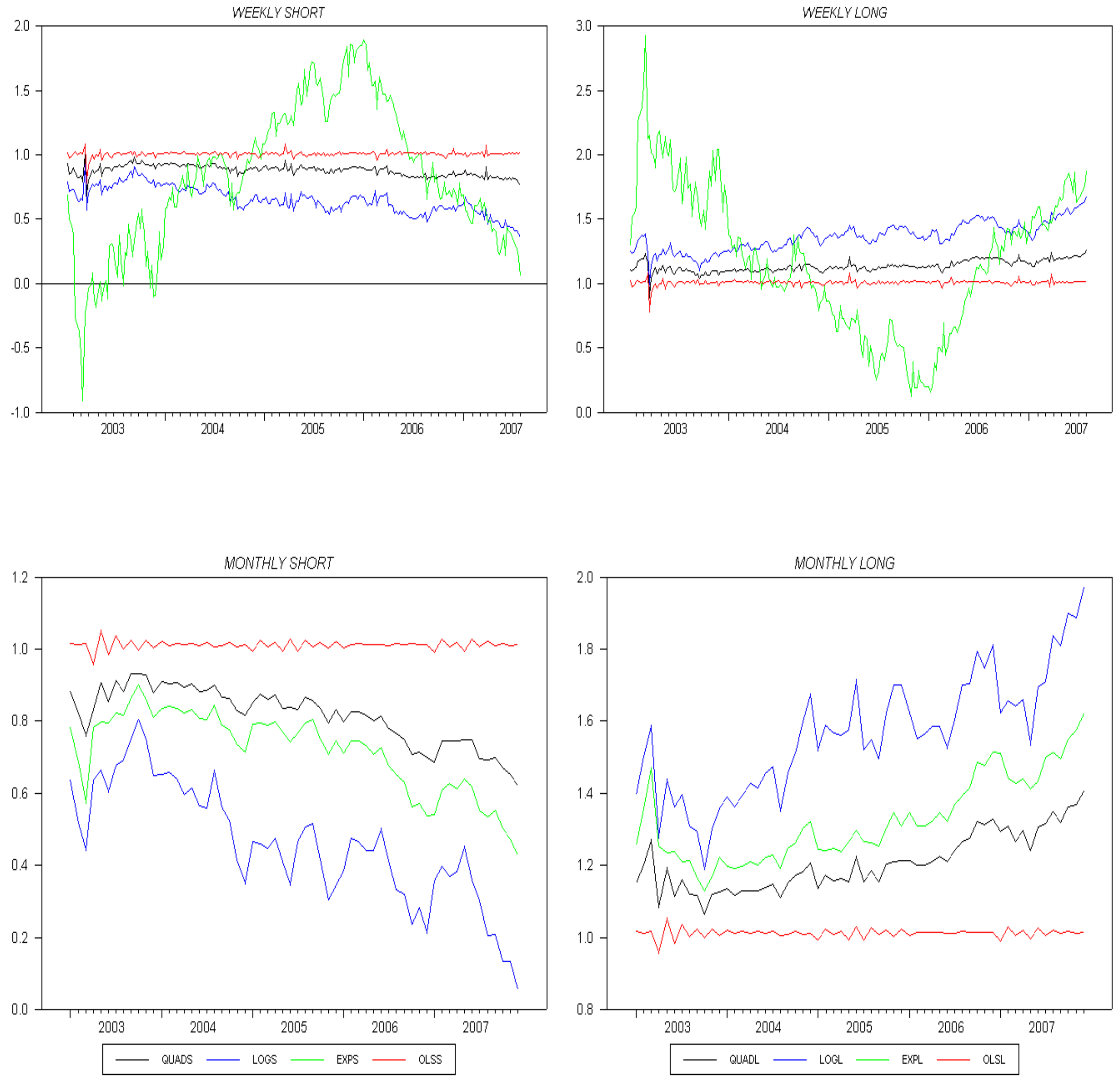

**Fig 3b: Time-varying Optimal Hedge Ratios for the Oil contract**

This figure is a time-series plot of the OHR's for the in-sample period for each of the different utility functions, Quadratic, Log, and Exponential together with the MVHR OHR for both short and long hedgers for both weekly and monthly hedging frequencies. The variance covariance matrix is calculated using a rolling window to allow each of the hedge strategies to vary over time. This allows us to compare hedges on the basis of utility as the variance covariance matrix underlying the optimal hedges is the same for each utility.